\documentstyle[12pt]{article}
\begin{document}

\setlength{\parskip}{.20in}

\begin{titlepage}

\title{A NOTE ON CLARK'S CONJECTURE ON TIME-WARPED BANDLIMITED SIGNALS\footnote{The author is with the Department of Electrical and Computer Engineering, University of Delaware, Newark, DE 19716, USA. Email: xxia@ece.udel.edu.}}

\author{Xiang-Gen Xia}

\date{}

\maketitle
\vspace{.2in}
\begin{center}
\underline{\bf Abstract}
\end{center}

In this note, a result of  a previous paper 
on  Clark's conjecture on time-warped
 bandlimited signals 
is extended to a more general class of the time warping functions
 $\gamma$, which includes most of the common functions in practice.

\end{titlepage}

\newpage
\pagenumbering{arabic}

\section{Introduction}

A bandlimited signal $f(t)$ can be reconstructed from its countable samples 
$f(t_{k})$ with certain sampling distance has resulted in many applications
 in digital signal processing and digital communication theory. In recent
 literature [1, 2, 3, 5], similar reconstructions for a non-bandlimited 
signal from its non-uniformly spaced samples have also
 been extensively discussed,
 which has a variety of applications in the above areas. In particular,
J. J. Clark et. al. [1, 2, 3] have introduced a  non-uniform sampling for 
the following time-warped bandlimited signals, which can be viewed as 
a generalization of the standard Shannon-Whittaker sampling theorem.

  Let $B$ denote the set of all finite positive energy real-valued 
signals defined in 
$(-\infty ,\infty)$ with spectra $F$ that vanish outside bounded intervals 
$[-\Omega _{f}, \Omega _{f}]$, and $\Gamma$ denote the collection of 
all real-valued monotone functions defined in $(-\infty , \infty )$. 
Then signals in $B\circ \Gamma = \{ f\circ \gamma$  formed by composing $f$ with 
$\gamma $ for $f\in B$ and $\gamma \in \Gamma  \}$ are called as 
time-warped bandlimited signals.

Notice that, for any $f\in B$, $f\not\equiv c$, where $c\in {\bf C}$ is 
an arbitrary constant.

If we consider $f\circ \gamma$ as a generalization of the standard phase
 modulation, the uniqueness (in certain sense) of the representation 
$(f \circ \gamma )(t)$ up to the scaling of $t$ is important
 for the demodulation process like what was mentioned in [2-5]. For this
 purpose, 
 in [2, 3] J. J. Clark stated the following conjecture.

\underline{\bf Conjecture 1}    If $f\in B$ and $\gamma \in \Gamma $ ,
then $f\circ \gamma \in B$ if and only if $\gamma $ is affine 
(i.e. $\gamma (t)=at+b$ with $a\neq 0$).

For more details of the representational ambiguity, one can see [2-5],
 in particular, Corollary 5.13, pp.83, [5].

In the approaches in [1-3], the invertibility of the time warping
$\gamma(t):{\bf R}\rightarrow {\bf R}$ (onto) is essential. It is known that
 a monotone 
function is invertible from ${\bf R}$ onto ${\bf R}$ if and only if
it is bijective, strictly monotone and continuous. Therefore, in what
follows, we only consider a time warping function 
$\gamma\in \Gamma_{0}$, where
\[\Gamma_{0}=\{\gamma:\,\gamma\,{\rm is\,continuous,\,
strictly\,monotone\,and\,bijective\,from}\,\,{\bf R}
\,\,{\rm to}\,\,{\bf R}\}\,.\]
Thus, 
\begin{equation}
\gamma\in \Gamma_{0} \Longleftrightarrow \gamma^{-1}\in \Gamma_{0}
\end{equation}
is clear.

About this conjecture, in [4] the following two results were obtained.

\underline{\bf Proposition 1}    If $f\in B$ and
 $\gamma \in \Gamma_{0} $ satisfies the condition that 
 $\gamma (t) = h(t)$ in some interval 
$[a,b]$ with $a<b$ for some $h(t)$ which is entire when $t$ is extended 
to ${\bf C}$, then, $f\circ \gamma \in B $  if and only if 
$\gamma $ is affine.

\underline{\bf Proposition 2}. If $f\in B$ and $\gamma\in \Gamma_{0}$
satisfies the condition that
$\gamma (t) = c_{1}t^{\alpha }+c_{0}$ in an interval $[a,b]$
with $a<b$ for some positive real number $\alpha $,  constants
$c_{1}\neq 0$ and $c_{0}$, then $f\circ \gamma \in B$
 if and only if
$\gamma $ is affine.

In this note, Proposition 1 is extended to a more general class of 
$\gamma$, which is that $\gamma(t)$(or $\gamma^{-1}(t)$)
$=h(t)$ in an interval $[a,b]$ for some 
$h(t)$ which can be extended to an analytic function in ${\bf C}$ 
except at possibly isolated points. Let
\[H=\{h(t):\,\,h(t)\,\,{\rm is\,analytic\,
except\,at\,possibly\,isolated\,points}\,\,{\rm when}\,\,t\,\,
{\rm is\,extended\,to}\,\,{\bf C}\}\,.\]
Obviously, the following functions are in $H$:
\[\sum_{k=1}^{m}\frac{b_{k}}{(t-a_{k})^{m_{k}}}+\sum_{k=1}^{n}c_{k}t^{n_{k}}
\,,\]
where $a_{k},b_{k}, c_{k}$ are arbitrary complex numbers and 
$m,n,m_{k},n_{k}$ are arbitrary integers. $H$ includes
 all entire functions. We now have the following new result on the conjecture.

\underline{\bf Theorem 1}. If  $f\in B$ and $\gamma \in \Gamma_{0}$
 satisfies
 the condition that  
 $\gamma (t)$(or $\gamma^{-1}(t)$)$= h(t)$ in some interval  
$[a,b]$ with $a<b$ for some $h(t)\in H$, 
then, $f\circ \gamma \in B $  if and only if  
$\gamma $ is affine.

  The idea to prove Theorem 1 is the following: we first prove $h(t)$
 can be extended to an entire function when $f \circ \gamma \in B$;
then,  by the result in [4] or [5], $h(t)$ has to be affine; finally, 
the result is obtained by
proving $\gamma(t)\equiv h(t)$ in ${\bf R}$.  In [4], a proof of 
$\gamma(t)\equiv h(t)$  
was missing. Lemma 2 in the next section gives the proof.

Let $H_{1}=\{$ any finite combinations of
additions, multiplications and divisions of entire functions$\}$ and 
$H_{2}=\{f:\,\,f\in H_{1}$ or $f^{-1}\in H_{1}\}$. Then $H_{2}$ includes
 most of the functions we can see in practice
  and the following Corollary 1 is straightforward.

\underline{\bf Corollary 1}. If $f\in B$ and $\gamma\in \Gamma_{0}$
satisfies the condition that $\gamma(t)=h(t)$ in some interval
$[a,b]$ with $a<b$ for some function $h(t)\in H_{2}$, then
$f\circ \gamma\in B$ if and only if $\gamma$ is affine. $\Box$

The result of Corollary 1 is practically satisfactory, because
we usually use functions in $H_{2}$ as the parts of the time  warping
functions $\gamma(t)$. 

{\bf Note}. A function $\gamma\in \Gamma$ is said to be strictly monotone
if
 $\forall t_{1},t_{2}\in {\bf R}$, if $t_{1}<t_{2}$
then $\gamma(t_{1})< \gamma(t_{2})$, or $\forall t_{1},t_{2}\in {\bf R}$,
if $t_{1}<t_{2}$ then $\gamma(t_{1})> \gamma(t_{2})$.

\section{Proof of Theorem 1}

Without any confusing in understanding, in what follows, the extended version
 of $g(t)$ to ${\bf C}$ is denoted by $g(z)$ for any 
possible function $g(t)$ defined
 in ${\bf R}$. Any $f\in B$ can be extended to an entire function of 
exponential type [6]. To prove Theorem 1, we need some lemmas.

\underline{\bf Lemma 1}.  Let $f\in B$ and $\gamma(t) \in \Gamma$
 satisfies the condition that $\gamma(t)=h(t)$ in some interval 
$[a,b]$ with $a<b$ for some $h(t)\in H$. If $f\circ \gamma \in B$, then
$h(t)\equiv \alpha t+\beta$ in ${\bf R}$ where $\alpha\neq 0$ and $\beta$ are real.

{\bf Proof}: $(f\circ \gamma)(z)$ is entire and $(f\circ h)(z)$ is analytic 
in ${\bf C}$ except at possibly isolated points. Since $(f\circ \gamma)(t)
=(f\circ h)(t)$ for $t\in [a,b]$, by the uniqueness of analytic functions
 $(f\circ \gamma)(z)=(f\circ h)(z)$ in ${\bf C}$ except at possibly isolated
points. But $(f\circ \gamma)(z)$ is entire. Therefore, all singularities
 of $f\circ h$ are removable. Thus, $(f\circ \gamma)(z)\equiv (f\circ h)(z)$ 
in ${\bf C}$. This implies that $(f\circ h)(t) \in B$.

To prove Lemma 1, we only need to prove that $h(t)$ can be extended to an 
entire function by the result in [4] or [5]. To do so,
first we prove that $h(z)$ has no essential singularities. 
Suppose $z_{0}\in {\bf C}$ is an essential singularity. Then Picard's
theorem [7] tells us that $h(z)$ can reach all values of ${\bf C}$ except 
possibly one when $z$ runs over a neighborhood of $z_{0}$ in ${\bf C}$. 
Since $f(z)$ is entire, $|f(z)|$ is unbounded when $z$ goes to infinity.
Therefore, $|(f\circ h)(z)|$ is also unbounded when $z$ runs over a neighborhood
 of $z_{0}$. It contradicts with the fact that $f\circ h$ is entire. This proves
 that $h(z)$ has no essential singularities. Next, we prove that $h(z)$ has no 
poles. Suppose $z_{0}\in {\bf C}$ is a pole of $h(z)$. Let 
\[h_{1}(z)=\frac{1}{h(z)}\,.\]
Then $z_{0}$ is a zero of $h_{1}(z)$ and $h_{1}(z)$ is analytic in a neighborhood
 of $z_{0}$. Therefore, there is a neighborhood $D(z_{0})$ of $z_{0}$ such that
$h_{1}(D(z_{0}))$ is an open set containing $0$ by the opening mapping 
theorem. This implies that $h(D(z_{0}))$ is an open set containing $\infty$.
The same as before, $|f\circ h|$ is not bounded in $D(z_{0})$, which contradicts
 with $f\circ h$ being entire. So, $h(z)$ has no poles. 
Since $h(z)$ has no other singularities
except isolated ones, $h(z)$
 has to be entire [7]. 

{\bf Remark}: A similar proof of $h(z)$ having no 
essential singularities or poles can be found in [5], pp.78-79, the proofs
of Lemma 5.3 
and Theorem 5.4.  $\Box$

\underline{\bf Lemma 2}. $f$, $\gamma$ and $\alpha t+\beta$ are as in
Lemma 1. If $f(\gamma(t))\equiv f(\alpha t+\beta)$ in ${\bf R}$, then
$\gamma(t)\equiv \alpha t+\beta$ in ${\bf R}$.

{\bf Proof}: Denote $D\stackrel{\Delta}{=}
\{t\in {\bf R}:\,\,f'(t)=0\}$. Since $f(z)$ is entire, $D$ is at most 
 a countable set and does not have any limit point in ${\bf R}$. Otherwise
 $f(z)$ is a constant, which contradicts with $f\in B$. 

Since points in $D$ are isolated, $D$ has only the following three forms.

i).
$D=\{\gamma_{i}:\,\,\gamma_{i}<\gamma_{i+1}$ for
$i=1,2,...,n\}$ for some positive integer $n$.

ii.
 $D=\{\gamma_{i}:\,\,\gamma_{i-1}<\gamma_{i}<+\infty$
 for any integer $i\leq 0$ and $\gamma_{n}\rightarrow -\infty$
 when $n \rightarrow -\infty$;
 or $-\infty<\gamma_{i}<\gamma_{i+1}$ for any integer $i\geq 0$ and $\gamma_{n}
\rightarrow +\infty$ when $n\rightarrow +\infty\}$. 

iii). $D=\{\gamma_{i}:\,\,\gamma_{i}<\gamma_{i+1}$ for any integer $i$, and 
$\gamma_{-n}\rightarrow -\infty$, $\gamma_{n}
\rightarrow +\infty$ when $n\rightarrow +\infty\}$

 Without loss
of generality, we may only consider the third form of $D$ and $\gamma$
being strictly increasing in the following proof.
 For the first form of $D$ or the decreasing 
$\gamma$, the proof is the same. Since $\gamma(t)=\alpha t+\beta$ for
$t\in [a,b]$, $\alpha>0$ for increasing $\gamma$. 

Since $f(t)$ is real,
$f(t)$ is strictly monotone in each interval $(\gamma_{i},\gamma_{i+1})$.

Let $t_{i}'=\gamma^{-1}(\gamma_{i})$ and $t_{i}''=(\gamma_{i}-\beta)/\alpha$
for any integer $i$. Then $t_{i}'<t_{i+1}'$ and $t_{i}''<t_{i+1}''$ for any
integer $i$, and 
\[\bigcup_{i=-\infty}^{\infty}[t_{i}',t_{i+1}']
=\bigcup_{i=-\infty}^{\infty}[t_{i}'',t_{i+1}'']={\bf R}\,.\]
Let $\bar{D}\stackrel{\Delta}{=}\{t_{i}':\,\,i=0,\pm 1,\pm 2,...\}
\bigcup \{t_{i}'':\,\,i=0,\pm 1,\pm 2,...\}$\\
$=\{t_{i}:\,\,t_{i}<t_{i+1}$ for any integer $i$, and $t_{-n}
\rightarrow -\infty$, $t_{n}\rightarrow +\infty$ when $n\rightarrow 
+\infty \}$. Then,
\[\bigcup_{i=-\infty}^{\infty}[t_{i},t_{i+1}]={\bf R}\,.\]
Suppose that, for integer $i_{0}$, $A_{0}\stackrel{\Delta}{=}
[a,b]\cap (t_{i_{0}},t_{i_{0}+1})\neq \phi$. Then $A_{0}$ is an interval
and $\gamma(t)=\alpha t+\beta,\forall\,t\in A_{0}$. Let 
$\tilde{t}_{i_{0}}\stackrel{\Delta}{=}inf\{t:\,\,t\in A_{0}\}$
and $\hat{t}_{i_{0}}\stackrel{\Delta}{=}sup\{t:\,\,t\in A_{0}\}$.
By the continuities of $\gamma(t)$ and $\alpha t+\beta$, 
\begin{equation}
\gamma(t)=\alpha t+\beta,\,\,\forall\,t\in [\tilde{t}_{i_{0}},
\hat{t}_{i_{0}}]\,.
\end{equation}

Let $t_{l}=sup\{t\leq \tilde{t}_{t_{0}}:\,\,\gamma(t)\neq 
\alpha t+\beta \}$ and $t_{r}=inf\{t\geq \hat{t}_{t_{0}}:
\,\,\gamma(t)\neq \alpha t+\beta \}$. 

To prove Lemma 2, by (2) we only need to
 prove that $t_{l}=-\infty$ and $t_{r}=+\infty$. Suppose $t_{r}<+\infty$,
 that is, $\hat{t}_{t_{0}}\leq t_{r}<+\infty$. Then, 
By the continuities of
$\gamma(t)$ and $\alpha t+\beta$,
\begin{equation}
\gamma(t)=\alpha t+\beta,\,\,\forall\,t\in [\tilde{t}_{i_{0}},
t_{r}]\,.
\end{equation}
 For $t_{r}$, there are the
following two cases.

First, if $f'(\gamma(t_{r}))\neq 0$, then $\gamma(t_{r})\in 
(\gamma_{i_{1}},\gamma_{i_{1}+1})$ for some integer $i_{1}$. In this
case, by the continuities of $\gamma(t)$ and $\alpha t+\beta$, there exists
$\delta(t_{r})>0$ such that $\gamma(t), \alpha t+\beta \in
(\gamma_{i_{1}},\gamma_{i_{1}+1}),\,\forall\,t\in [t_{r},
t_{r}+\delta(t_{r})]$.
 Since $f(t)$ is strictly
 monotone in $(\gamma_{i_{1}},\gamma_{i_{1}+1})$, $f(\gamma(t))
\neq f(\alpha t+\beta)$ whenever $\gamma(t)\neq \alpha t+\beta$ for
$t\in [t_{r},t_{r}+\delta(t_{r})]$.
But $f(\gamma(t))\equiv f(\alpha t+\beta)$ in ${\bf R}$. Therefore,
\begin{equation}
\gamma(t)=\alpha t+\beta,\,\,\forall\,t\in [t_{r},
t_{r}+\delta(t_{r})]\,.
\end{equation}
By (3), (4) and the definition of $t_{r}$, $t_{r}\geq t_{r}+\delta(t_{r})$. 
This is a contradiction. 

Second, if $f'(\gamma(t_{r}))=0$, then $\gamma(t_{r})=\gamma_{i_{1}}$ 
for some integer $i_{1}$.
 Since $\gamma(t)$ and
$\alpha t+\beta$ are both increasing and continuous, there exists
$\delta(t_{r})>0$ such that $\gamma(t),\alpha t+\beta \in
(\gamma_{i_{1}},\gamma_{i_{1}+1}),\,\,\forall\,t\in
(t_{r},t_{r}+\delta(t_{r})]$. Similar to the first case, (4) also holds and
$t_{r}\geq t_{r}+\delta(t_{r})$. This is also a contradiction.

Thus, $t_{r}=+\infty$ is proved. $t_{l}=-\infty$ can be proved similarly.
Therefore, $\gamma(t)\equiv \alpha t+\beta$ in ${\bf R}$. $\Box$

{\bf Proof of Theorem 1}:

We only need to prove the ``only if'' part.

If $\gamma(t)=h(t)$ in some interval $[a,b]$ with $a<b$ for 
some $h(t)\in H$, then Theorem 1 is a consequence of Lemma 1
 and Lemma 2. 

If $\gamma^{-1}(t)=h(t)$ in some interval $[a,b]$ with $a<b$ for
some $h(t)\in H$, we let $g(t)\stackrel{\Delta}{=}
f(\gamma(t))\in B$ and $\gamma_{1}\stackrel{\Delta}{=}
\gamma^{-1}$. Then, $g(\gamma_{1}(t))=f(t)\in B$. For $\gamma_{1}$,
by (1) and the above proof, $\gamma_{1}(t)\equiv \alpha t+\beta$ in ${\bf R}$
  for some real 
$\alpha\neq 0$ and $\beta$. Therefore, $\gamma(t)\equiv \frac{1}{\alpha}
t-\beta$ in ${\bf R}$. Theorem 1 is proved.
$\Box$

\section{Conclusion}
This paper extended a result on Clark's conjecture on time-warped bandlimited
signals. It has been recently found applications in [8].

This paper was originally written in 1990.

\begin{center}
{\bf REFERENCES}
\end{center}

[1] J. J. Clark, M. R. Palmer, and P. D. Lawrence, ``A transformation method for the reconstruction of functions from nonuniformly spaced samples,'' {\em IEEE Trans. on Acoustics, Speech, and Signal Processing}, ASSP-33(4), pp.1151-1165, Oct. 1985.

[2] D. Cochran and J. J. Clark, ``On the sampling and reconstruction of time-warped bandlimited signals,'' {\em Proceeding of ICASSP'90}, Vol. 3, pp.1539-1541, April 1990.

[3] J. J. Clark, "Sampling and reconstruction of bandlimited signals," {\em Proceedings of Visual Communications and Image Processing IV }, SPIE Vol. 1199, pp. 1556-1562, 1989.

[4] X.-G. Xia and Z. Zhang, ``On a conjecture on time-warped bandlimited signals,''
  {\em IEEE Trans. on Signal Processing}, vol. 40, no. 1, pp. 252-254, Jan. 1992.

[5] D. Cochran, {\em Nonlinear Problems in Signal Analysis}, Ph.D. thesis, Harvard University, 1990. 

[6] P. H. Boas, {\em Entire Functions}, Academic Press Inc., 1954.

[7] W. Rudin, {\em Real and Complex Analysis}, McGraw-Hill, 1974.

[8] H. Liu, X.-G. Xia, and R. Tao,
``Variation of a signal in Schwarzschild spacetime,''
{\em Science China--Information Sciences}, vol. 62, no. 8,
82304, Aug. 2019.\\
https://doi.org/10.1007/s11432-019-9856-y

\end{document}